\documentclass[amsmath,amssymb,aps,prl,reprint,superscriptaddress]{revtex4-1}
\pdfoutput=1

\usepackage{mathptmx}

\usepackage[squaren]{SIunits}  
\usepackage{graphicx}    
\usepackage{color}

\usepackage{hyperref}

\newcommand{\bra}[1]{\langle #1|}
\newcommand{\ket}[1]{|#1\nolinebreak[4]\rangle}

\newcommand{\abs}[1]{\left\lvert #1 \right\rvert}

\DeclareMathOperator{\EF}{EF}

\begin{document}

\title{Quantum entanglement distribution using a magnetic field sensor}

\author{Marcus Schaffry}
\affiliation{Department of Materials, University of Oxford, Parks Road, Oxford OX1 3PH, United Kingdom}
\author{Simon C. Benjamin}
\affiliation{Department of Materials, University of Oxford, Parks Road, Oxford OX1 3PH, United Kingdom}
\affiliation{Centre for Quantum Technologies, National University of Singapore, 3 Science Drive 2, Singapore 117543}
\author{Yuichiro Matsuzaki}
\affiliation{Department of Materials, University of Oxford, Parks Road, Oxford OX1 3PH, United Kingdom}

\date{\today}

\begin{abstract}
Sensors based on crystal defects, especially nitrogen vacancy (NV) centres in nanodiamond, can achieve detection of single magnetic moments. Here we show that this exquisite control can be utilized to entangle remote electronic spins for applications in quantum computing; the mobile sensor provides a `flying' qubit while the act of sensing the local field constitutes a two-qubit projective measurement. Thus the tip mediates entanglement between an array of well-separated (and thus well controlled) qubits. Our calculations establish that such a device would be remarkably robust against realistic issues such as dephasing and multimodal vibrations in the sensor tip. We also provide calculations establishing the feasibility of performing a demonstrator experiment with a fixed sensor in the  immediate future.
\end{abstract}

\maketitle

One possible architecture for a quantum computer is based on the idea of distributed quantum information processing (QIP) \cite{Cirac.Ekert.ea1999Distributedquantumcomputation,Lim.Beige.ea2005Repeat-Until-SuccessLinearOptics,Benjamin.Browne.ea2006Brokeredgraph-statequantum}. Here, rather than controlling all qubits on one localized site the set of qubits is distributed to various spatially separated sites that are entangled with each other. This physical distance confers the benefit of better control of the individual qubits. Typically for the entanglement operation, an optical setup with photons is proposed \cite{Barrett.Kok2005Efficienthigh-fidelityquantum}. In this Letter, however we show how one can use the dipole-dipole interaction between electronic spins in conjunction with optical detected magnetic resonance (ODMR) to create entanglement between different sites. The NV centre defects in diamond are very suitable for ODMR and QIP as these possess a long-lived spin triplet electronic ground state with the levels $\ket{0}$ and $\ket{\pm1}$ that can be easily initialized with a laser, manipulated with microwave pulses and read-out optically \cite{Wrachtrup.Kilin.ea2001Quantumcomputationusing,Jelezko.Gaebel.ea2004ObservationofCoherent}. This exquisite control enables observation of their coupling to adjacent nuclear spins \cite{Gaebel.others2006Room-temperaturecoherentcoupling,Dutt.others2007QuantumRegisterBased} and the measurement of a nearby nuclear spin \cite{Neumann.others2010Single-ShotReadoutof}. A very promising application of NV centres is their capability to detect the strength of very small magnetic fields through an induced Zeeman splitting \cite{Taylor.Cappellaro.ea2008High-sensitivitydiamondmagnetometer,Maze.Stanwix.ea2008Nanoscalemagneticsensing,Degen2008Scanningmagneticfield,Balasubramanian.Chan.ea2008Nanoscaleimagingmagnetometry,Schaffry.others2011SpinAmplificationMagnetic}. In the following we will show how this high sensitivity to magnetic fields can be used to entangle two remote electronic spins.

\begin{figure}[hbt!]
  \centering
 \includegraphics[width=1\linewidth]{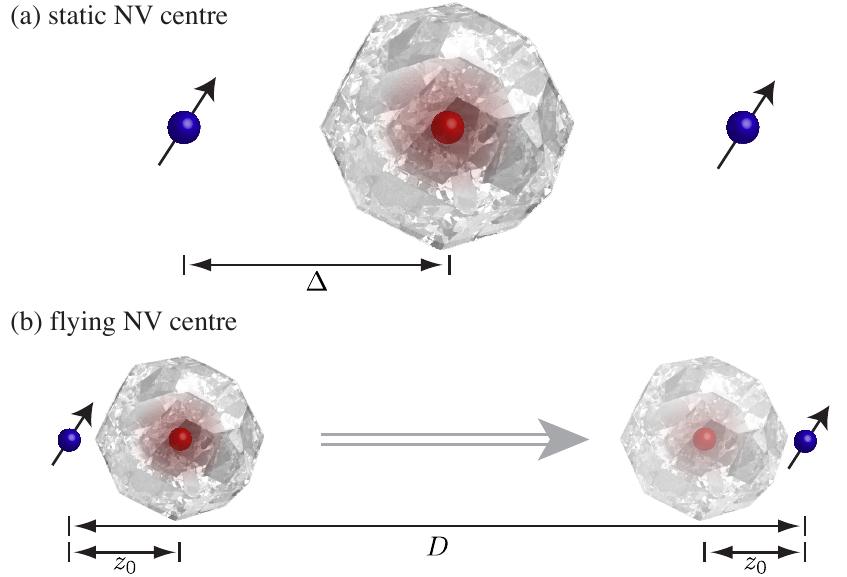}
 \caption{(a) An NV centre in a nanocrystal is placed in between two adjacent electronic spins qubits. The dipole-dipole interaction between the spins and a measurement of the NV centre allows to entangle the two qubits. (b) An NV centre is initially close to one qubit and then it is moved near to the second qubit that is far apart from the first qubit. Measuring the NV centre at the location close to the second qubit again allows entanglement of the two qubits.}
  \label{fig:schematic}
\end{figure}

Ultimately our proposal is to move a NV centre sensor between remote spins to entangle them. We begin by outlining a more simple experimental scenario, which could be tested in the immediate future. Suppose we are given two electronic spin qubits and we can measure the field of these two spins by using a crystal defect in a nanodiamond which is placed in the middle of the two qubits (see Fig.~\ref{fig:schematic}a). If one qubit produces a field of strength $b_z$ at the site of the NV centre then the NV centre experiences either $-2b_z,0,$ or $2b_z$ depending on the spin orientation of the qubits: the field is $0$ if the two spins are in the state $\ket{\uparrow\downarrow}$ or $\ket{\downarrow\uparrow}$ and $\pm 2b_z$ if the two spins are in the state $\ket{\downarrow\downarrow}$ or $\ket{\uparrow\uparrow}$. Given a sufficiently large external magnetic field, we can prepare the NV centre in the state $\ket{+}_{\text{NV}}=1/\sqrt{2}\bigl(\ket{0}+\ket{1}\bigr)$. Over time $t$, this state collects either a phase of $\pm2b_z \mu_{\text{NV}}t/\hbar\equiv \omega_{\pm}t$ or $0$ depending on the spin state of the qubits, where $\mu_{\text{NV}}$ denotes the magnetic moment of the NV centre. Hence this phase allows us to perform a projective two qubit measurement. For example, suppose the two qubits are initially also prepared in a $\ket{+}_{1/2}=1/\sqrt{2}\bigl(\ket{\downarrow}+\ket{\uparrow}\bigr)$ state and we let the NV centre precess for the time $\tau=\pi/\omega_+$, before measuring it in the $\ket{\pm}_{\text{NV}}=1/\sqrt{2}\bigl(\ket{0}\pm\ket{1}\bigr)$-basis. If the measurement results in $\ket{+}_{\text{NV}}$, then the two qubits will be in the Bell state $1/\sqrt{2}\bigl(\ket{\downarrow\uparrow}+\ket{\uparrow\downarrow}\bigr)$ and similarly, if the outcome is $\ket{-}_{\text{NV}}$ then the two qubits will be in the Bell state $1/\sqrt{2}\bigl(\ket{\downarrow\downarrow}+\ket{\uparrow\uparrow}\bigr)$. We therefore implemented a deterministic parity projection. As we will show, for times other than the ideal $\tau$ the procedure still achieves a probabilistic parity projection. Such a projection is known to be sufficient to implement the entanglement needed for full QIP \cite{Barrett.Kok2005Efficienthigh-fidelityquantum}.

\begin{figure}[hbt]
  \centering
\includegraphics[width=1\linewidth]{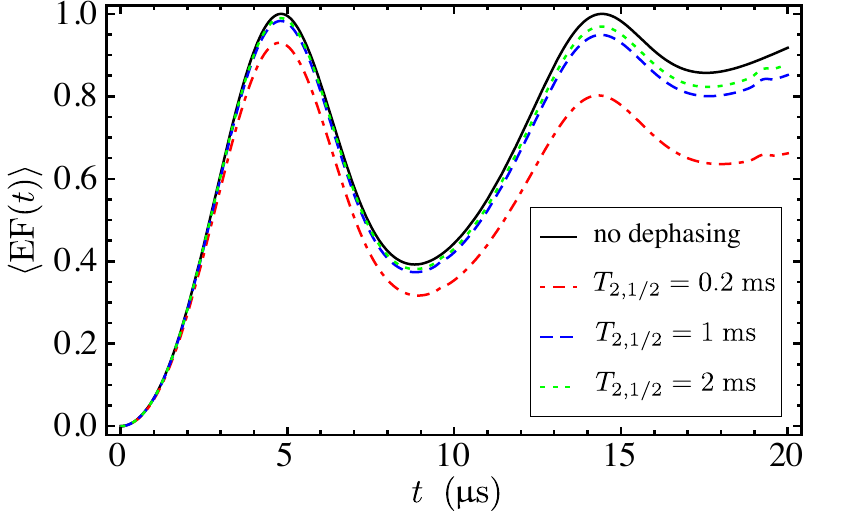}
\caption{(Color online) Static NV centre: mean EF with respect to the measurement time. For the non-solid curves: $T_{2,\text{NV}}=\unit{2}{\milli\second}$. The distance between the NV centre and the qubits is $\Delta=\unit{10}{\nano\meter}$.}
  \label{fig:static-EF}
\end{figure}

We now analyse this general idea in a more rigorous way to show how robust it is with respect to the translations and vibrations of the NV centre. First, we look at a configuration where the two qubits are located at the origin and at $z=2\Delta$ and the NV centre which is oriented along the $z$-direction, is placed in the middle of the two qubits (see Fig.~\ref{fig:schematic}a)). The interaction between the three particles is given by the dipole-dipole coupling:
\begin{equation}
\label{eq:dd-interaction}
  H_{D,ij}  = C_{i,j}d_{i,j}^{-3} \bigl( 3 (\hat{\mathbf{x}}_{i,j}\cdot \mathbf{S}_i) (\hat{\mathbf{x}}_{i,j}\cdot \mathbf{S}_j) - \mathbf{S}_i\cdot\mathbf{S}_j \bigr) \quad;
\end{equation}
where $\mathbf{S}_i$ is the spin-operator of the particle $i$, $\hat{\mathbf{x}}_{i,j}$ is a unit vector pointing from spin $i$ to spin $j$, $d_{i,j}$ is the physical distance between the spins $i$ and $j$ and
$C_{i,j}=-\frac{\mu_0}{4\pi} \mu_i \mu_j=C$ is a constant \cite{Levitt2008SpinDynamics}. Here $\mu_0$ is the magnetic constant, $\mu_{1/2/\text{NV}}=2\mu_B$ are the magnetic moments of the spins, and $\mu_B$ denotes the Bohr magneton. In an external magnetic field $B$ in $z$-direction, the whole system can be described by the following Hamiltonian:
\begin{align}
  H_{\text{static}} &=H_0+H_{\text{DIP}} \quad \text{with}\\
 H_0&= -\mu_{\text{NV}} BS_{z,\text{NV}} + D_{\text{NV}} S_{z,\text{NV}}^2-\mu_1 BS_{z,1}-\mu_2 BS_{z,2}\\
  H_{\text{DIP}}&= H_{D,\text{NV}1} + H_{D,\text{NV}2}+H_{D,12}  \quad,
\end{align}
here $D_{\text{NV}}=\unit{2.87}{\giga\hertz}$ is the zero-field splitting (ZFS). We transform this Hamiltonian to a rotating frame with respect to $\exp(iH_{0}t)$ and neglect fast oscillating terms originating from a sufficiently large external field and the ZFS (rotating wave approximation). This gives us:
\begin{align}
\tilde{H}_{\text{static},\text{RWA}}  &=H_{D,\text{app},\text{NV}1}+H_{D,\text{app},\text{NV}2}+H_{D,12} \;, \text{with}    \label{eq:static-RWA-Hamiltonian} \\
 H_{D,\text{app},ij} &= 2Cd_{i,j}^{-3}  S_{z,i} S_{z,j}  \quad. \label{eq:dd-interaction-approx}
\end{align}
By using a standard master equation \cite{Breuer.Petruccione2002TheoryofOpen} as follows we can now evaluate the negative effect caused by dephasing upon our basic proposal:
\begin{multline}
  \rho'(t)= -i [ \tilde{H}_{\text{static},\text{RWA}} , \rho(t)]  \\+\sum_{j={1,2,\text{NV}}} \frac{2}{T_{2,j}} \Bigl( S_{j,z} \rho(t) S_{j,z}^{\dagger} -\frac{1}{2} \{ \rho(t), S_{j,z}^{\dagger} S_{j,z} \} \Bigr) \quad.
\end{multline}
We assume the three particles to be initially in the state
\begin{equation}
\label{eq:initial-state}
  \ket{\psi_i} =\ket{\psi(0)}= \ket{+}_1 \ket{+}_{\text{NV}} \ket{+}_2  \quad.
\end{equation}
After time $t$ we measure the NV centre in the $\ket{\pm}_{\text{NV}}$-basis. The qubits are then, depending on the outcome, either in the state $\rho_+$ or $\rho_-$. This holds for both with the probability
\begin{equation}
\label{eq:prob-static}
  p_{\pm} = \bigl( 1 \pm \exp(-t/T_{2,\text{NV}})\cos(\alpha t/2)^2 \bigr)/2 \quad,
\end{equation}
where $\alpha=2C\Delta^{-3}$. Within the limits of infinite dephasing times $\rho_{\pm}=\ket{\psi_{\pm}}\bra{\psi_{\pm}}$ are pure where
\begin{align}
\ket{\psi_+}&=n_+ \Bigl(e^{-i\alpha t} \ket{\downarrow\downarrow}+\ket{\uparrow\uparrow}+\frac{2e^{i \frac{3\alpha}{32} t}}{1+e^{i\alpha t}}(\ket{\downarrow\uparrow}+\ket{\uparrow\downarrow})\Bigr) \nonumber \\
\ket{\psi_-}&=1 /\sqrt{2} (-e^{-i\alpha t} \ket{\downarrow\downarrow}+\ket{\uparrow\uparrow}) \nonumber
\end{align}
and where $n_+$ is a normalization factor. Our entanglement operation depends on the measurement outcome. Hence in order to quantify the amount of entanglement between the qubits that we obtain after measuring the NV centre at time $t$, we calculate the mean entanglement of formation (EF) \cite{Wootters1998EntanglementofFormation}, i.e.\
\begin{equation}
  \langle \EF(t) \rangle := p_- \EF(\rho_-) +p_+ \EF(\rho_+)  \quad,
\end{equation}
which we then plot with respect to the time of the measurement in Fig.~\ref{fig:static-EF}. Note that for QIP we could simply discard $\rho_+$ retaining $\rho_-$ which represents a perfect parity projection if we know $t$.
In this plot we confirm the expected oscillatory behaviour due to the phase that the NV centre collects. In addition one can show that the distortion of the oscillation is a result of the
qubit-qubit coupling in Eq.~\eqref{eq:static-RWA-Hamiltonian} and due to the relatively small magnitude of these terms that they have negligible effect on the first maximum of the mean EF. Finally we conclude that the effect of dephasing for the first maximum of the mean EF is also small when the coherence times are sufficiently large. We can expect large $T_2$ times if we implement the qubits for example with Sc@C${}_{82}$ or an NV centre.  For NV centres coherence times of about \unit{2}{\milli\second} have already been measured at room temperature \cite{Balasubramanian.others2009Ultralongspincoherence,Naydenov.others2011Dynamicaldecouplingof}. Additionally Sc@C${}_{82}$ has an unpaired electron spin that mainly exists on the fullerene cage \cite{Ge.others2008Modelingspininteractions} with a coherence time of \unit{200}{\micro\second} \cite{Morley.others2005Hyperfinestructureof, Brown.others2010Electronspincoherence}. Hence our entangling operation is capable of achieving high fidelity entanglement.

\begin{figure}[hbtp]
  \centering
  \includegraphics[width=1\linewidth]{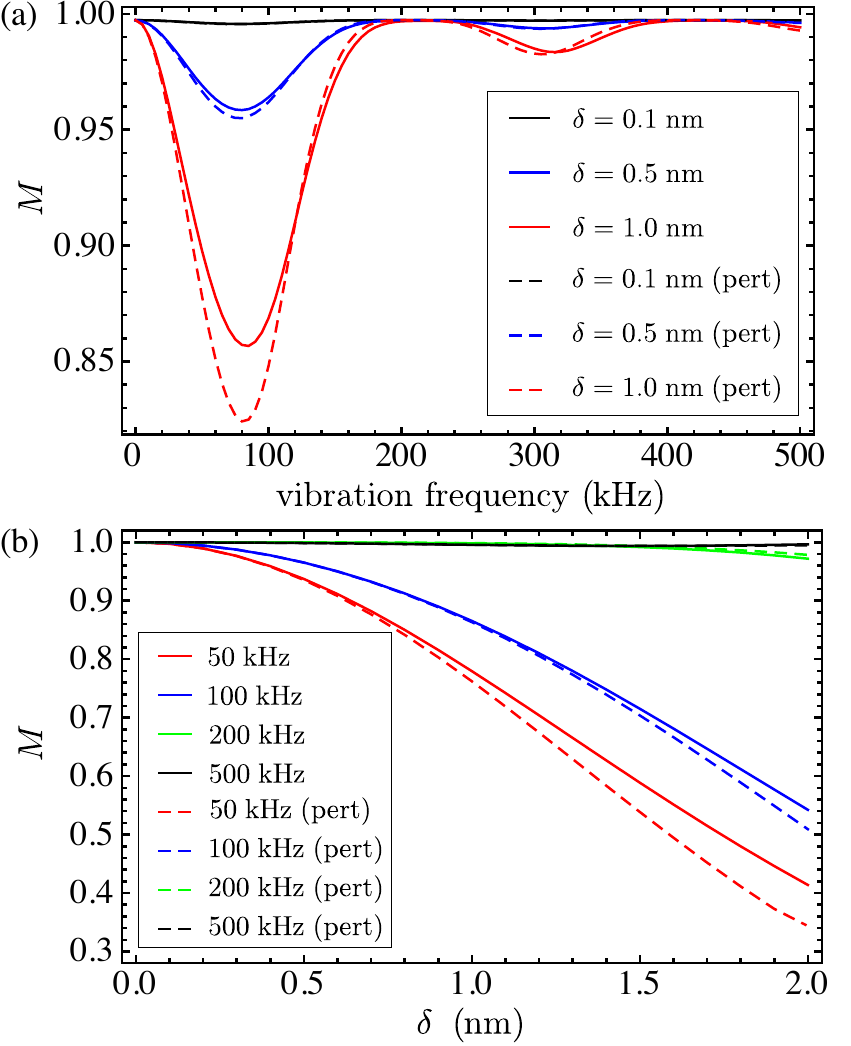}
 \caption{(Color online) (a) Maximal achievable mean EF for vibration modes where the vibration frequencies (amplitudes) are distributed according to a truncated normal distribution on $[0,\infty)$; with mean $\omega$ ($\delta$) and standard deviation $0.01\omega$ ($0.01\delta$), i.e.\ $\overline{\rho}^{\omega}$ ($\overline{\rho}^{\delta}$) with respect to the mean (common) vibration frequency $\omega/2\pi$ for different common (mean) vibration amplitudes $\delta$ and $\phi=\frac{\pi}{2}$. (b) Maximal achievable mean EF for vibration modes where the phase is uniformly distributed, i.e.\ $\overline{\rho}^{\phi}$ with respect to their common vibration amplitude $\delta$ for different common vibration frequencies. 
In both plots the solid lines are calculated from the numerical solution of the von Neumann equation for $\tilde{H}_{\text{vib,RWA}} (t)$ and the dashed lines are calculated by using the approximations discussed in the text. Here $\Delta=\unit{10}{\nano\meter}$.}
  \label{fig:vibration-avgs}
\end{figure}

Next, we address a more practical consideration. Namely that the nanocrystal in which the NV centre is embedded can be placed on an AFM-tip which then enables us to move the defect very precisely between various sites in a distributed quantum computer architecture. This ability of spatial control comes with the price of limited controllable vibrational modes of the AFM-tip, with typical frequencies varying between \unit{50}{\kilo\hertz} and \unit{500}{\kilo\hertz} \cite{Yun.Park.ea1999Regiospecificorientationof}. To characterize the consequences of these vibrations onto our entangling operation we consider first the effect of the NV centre oscillating between the two qubits in a single mode, i.e.\ $d_{\text{NV,1/2}}(t) = \Delta \pm \delta \cos(\omega t +\phi)$; where $\delta$ denotes the amplitude, $\omega$ the frequency, and $\phi$ the phase of the oscillation. As above, we derive a rotating wave approximation Hamiltonian and end up with the time-dependent Hamiltonian
\begin{equation}
\tilde{H}_{\text{vib},\text{RWA}} (t) =H_{D,\text{app},\text{NV}1}(t)+H_{D,\text{app},\text{NV}2}(t)+H_{D,12} \; . \label{eq:Hamiltonian-vib}
\end{equation}
We can consider this vibrational scenario as a perturbed static case and hence we expect a maximum of the mean EF at around $\pi/\abs{\alpha}$ (similar to Fig.~\ref{fig:static-EF}). Note that for larger times the qubit-qubit interaction and the dephasing become relevant. For this reason we define the maximal achievable mean EF $M$ to be
\begin{equation}
  M=M(\delta,\omega,\phi)=\max_{0\leq t \leq 2\pi/\abs{\alpha}} \langle \EF(t) \rangle \quad.
\end{equation}
We are interested in how the introduction of many modes with various parameters $(\delta,\omega,\phi)$ affect this maximal achievable mean EF, and in particular, for which parameter regime $M$ is close to $1$. To this purpose, we analyse the effect of many modes that share the common parameter $p$ on $M$ and define
\begin{equation}
\label{eq:rho-avg}
 \overline{\rho}^p(t)= \int_{-\infty}^{\infty} \mathcal{D}(p) \ket{\psi(t,p)}\bra{\psi(t,p)}dp \quad,
\end{equation}
as the mean density matrix with respect to the distribution $\mathcal{D}(p)$ of the parameter $p$; where $\ket{\psi(t,p)}$ is the solution of the Schr\"{o}dinger equation for $\tilde{H}_{\text{vib},\text{RWA}} (t)$. We find an analytic approximation of this solution as follows. First we neglect the qubit-qubit coupling in $\tilde{H}_{\text{vib},\text{RWA}} (t)$
\begin{equation}
\tilde{H}_{\text{vib},\text{RWA},\text{app}} (t) =H_{D,\text{app},\text{NV}1}(t)+H_{D,\text{app},\text{NV}2}(t) \quad. \label{eq:Hamiltonian-vib-app}
\end{equation}
We then transform this Hamiltonian to an interaction picture and expand it in a power series with respect to $\delta(t)=\delta \cos(\omega t +\phi)$. This gives:
\begin{align}
\label{eq:intPicHam}
&H_I(t)=  e^{i \tilde{H}_{\text{static},\text{RWA}} t}  \tilde{H}_{\text{vib},\text{RWA},\text{app}}  e^{-i \tilde{H}_{\text{static},\text{RWA}} t}\\ 
&=\sum_{j=1,2} \frac{2C}{\Delta^3}\left((-1)^j 3 \frac{\delta(t)}{\Delta}+6\frac{\delta(t)^2}{\Delta^2}  + O(\delta(t)^3) \right) S_{z,\text{NV}}S_{z,j} \;. \nonumber
\end{align}
Finally, we apply time-dependent perturbation theory:
\begin{multline}
\label{eq:app-sol}
  \rho_I(t) = \rho_I(0) - i \int_0^t [H_I(t'),\rho_I(0)] dt' \\
+ (-i)^2 \int_0^t \int_0^{t'} [H_I(t'),[H_I(t''),\rho_I(0)]]dt'' dt' \quad,
\end{multline}
where $\rho_I(0)=\ket{\psi_i}\bra{\psi_i}$ and get an analytic approximation of the solution of the Schr\"{o}dinger equation for Eq.~\eqref{eq:Hamiltonian-vib}. The first order approximation in $\delta$ reads:
\begin{multline}
\label{eq:rhoPerturbation2}
  \rho_I(t) = \rho_I(0) + i \frac{6C}{\Delta^4}\delta \frac{\sin(\omega t +\phi)-\sin(\phi)}{\omega} \\
\times [S_{z,\text{NV}}S_{z,1}-S_{z,\text{NV}}S_{z,2},\rho_I(0)] \quad.
\end{multline}
This approximation allows us to understand the effect of many modes. First, we analyse the effect of many vibrational modes with different frequencies. In Fig.~\ref{fig:vibration-avgs}a we compare our approximation with the exact solution for \eqref{eq:Hamiltonian-vib} by plotting the maximal achievable mean EF for $\overline{\rho}^{\omega}$ (in both cases) with respect to the vibration frequency $\omega$ for different amplitudes and for a particular phase. For the average $\overline{\rho}^{\omega}$ in Eq.~\eqref{eq:rho-avg} we use a truncated Normal distribution on $[0,\infty)$ with mean $\omega$ and standard deviation $\sigma=0.01\omega$. In the static case we have seen that the mean EF is maximal when we wait about time $\pi/\abs{\alpha}$ before we measure the NV centre. First assume the NV centre is initially in the middle of the two qubits ($\phi=\pi/2$). If the NV centre oscillates for $1/2+k$ periods ($k$ periods), where $k$ is a non-negative integer before it is measured, then the asymmetry in the system is maximal (minimal) and we get a low (high) mean EF. Obviously for a large $k$, this effect is less pronounced. The minus sign in the commutator in Eq.~\eqref{eq:rhoPerturbation2} describes exactly this asymmetry which decreases for large frequencies and thus the perturbative solution \eqref{eq:rhoPerturbation2} explains the characteristics of Fig.~\ref{fig:vibration-avgs}a. We note that analogous plots where we increase and decrease the width of the frequency distribution are not very sensitive to the width of the distribution which is in agreement with the rate in Eq.~\eqref{eq:rhoPerturbation2}.

Second we can repeat this analysis by averaging over the amplitude instead of the frequency, i.e.\ by considering $\overline{\rho}^{\delta}$. Again we see from Eq.~\eqref{eq:rho-avg} that averaging over a reasonably narrow truncated Normal distribution on $[0,\infty)$ has little effect, and hence, we get the same plot as in the analysis before. 

Third we analyse the effect of many vibrational modes with different phases. Thus we assume a uniform distribution of phases on $[0,2\pi]$. With Eq.~\eqref{eq:app-sol} we obtain in second order in $\delta$ the following density matrix 
\begin{multline}
\label{eq:solAVGphase}
  \overline{\rho}^{\phi}_I(t) = \rho_I(0) -i 6\frac{C}{\Delta^5}\delta^2 t [S_{z,\text{NV}}S_{z,1}+S_{z,\text{NV}}S_{z,2},\rho_I(0)]\\
- 18 \frac{C^2}{\Delta^8} \delta^2 \frac{1-\cos(\omega t)}{\omega^2}  \\
\times [S_{z,\text{NV}}S_{z,1}-S_{z,\text{NV}}S_{z,2},[S_{z,\text{NV}}S_{z,1}-S_{z,\text{NV}}S_{z,2},\rho_I(0)]] \;.
\end{multline}
Again we compare this approximation with the exact solution for \eqref{eq:Hamiltonian-vib} by plotting in Fig.~\ref{fig:vibration-avgs}b the maximal achievable mean EF with respect to the vibration amplitude for different frequencies and find a good agreement between the two solutions especially for small amplitudes. We see that the last term in \eqref{eq:solAVGphase} can be seen as a (time-dependent) decoherence rate and explains why higher frequencies and smaller amplitudes are less disturbing to our entangling operation.

In summary we conclude that if oscillations of the NV centre are not avoidable then we can mitigate the adverse effect for our entangling operation by ensuring that the oscillations are very fast and of low amplitude.

Having established that a static NV centre can entangle two external spins, and having further determined that the realistic issues of dephasing and oscillation do not present fundamental difficulties, we are now in a position to consider entangling two \emph{remote} spins via a moving, or 'flying', NV centre. Therefore we consider a distance $D\gg \Delta$ between the two qubits and propose that the NV centre should now fly from one qubit to the other, i.e.\ from $z=z_0$ to $z=D-z_0$ with velocity $v$. Hence the NV centre interacts first with qubit $1$ and then with qubit $2$ before it is measured at $t_M=(D-2z_0)/v$ (see Fig.~\ref{fig:schematic}b). 

As above we derive with the time-dependent distances
\begin{equation}
  d_{\text{NV},1}(t)=tv+z_0 \text{ and } d_{\text{NV},2}(t)= D-tv-z_0 \quad,
\end{equation}
the following rotating frame approximation Hamiltonian
\begin{equation}
\tilde{H}_{\text{flying,RWA}} (t) =H_{D,\text{app},\text{NV}1}(t)+H_{D,\text{app},\text{NV}2}(t) \quad,
\end{equation}
where we neglect the qubit-qubit coupling due to the large distance between the qubits.

\begin{figure}[h!]
  \centering
\includegraphics[width=1\linewidth]{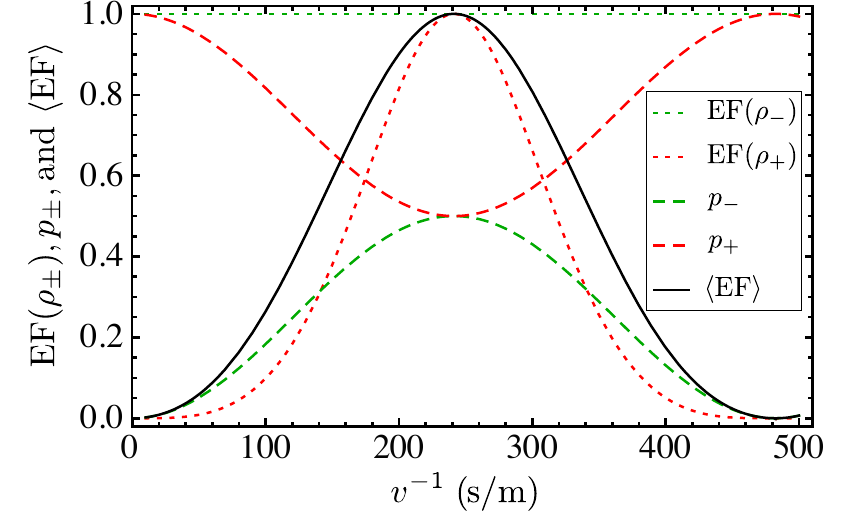}
 \caption{(Color online) (a) Flying NV centre: EFs, mean EF and success probabilities as a function of the inverse velocity $v^{-1}$ when the NV centre has moved from one qubit to the other. Parameters: $D=\unit{100}{\nano\meter},z_0=\unit{5}{\nano\meter}$.}
  \label{fig:flying-EF}
\end{figure}

Again we initialize the three spins in the state $\ket{\Psi_i}=\ket{\psi_i}$ (see Eq.~\eqref{eq:initial-state}). The Schr\"{o}dinger equation determines the state $\ket{\Psi(t_M)}$ when the NV centre travelled from one qubit to the other. The measurement of the NV centre in the $\ket{\pm}_{\text{NV}}$-basis projects the qubits either to the state
\begin{align}
\ket{\Psi_-}&=(-e^{i\beta}\ket{\downarrow\downarrow}+\ket{\uparrow\uparrow} )/\sqrt{2} \quad \text{or}\\
\ket{\Psi_+}&= N_+ \Bigl(e^{i\beta}\ket{\downarrow\downarrow}+\ket{\uparrow\uparrow} +\frac{2e^{i\beta}}{1+e^{i\beta}}(\ket{\downarrow\uparrow}+\ket{\uparrow\downarrow})\Bigr)
\end{align}
each with the probability:
\begin{align}
\label{eq:prob2}
 p_{\text{fly},-} = \sin^2(\beta/2 )/2\;\; \text{and} \;\;  p_{\text{fly},+} = ( 3+ \cos(\beta) )/4 \quad ,
\end{align}
where $\beta = C \left(  (D-z_0)^{-2}-z_0^{-2}\right)/v$ and $N_+$ is a normalization factor. The EF of the first state is $1$ and that of the second state is given by the binary entropy function $H$
\begin{equation}
   \EF(\rho_+)=H\Bigl(\frac{1}{2} + \frac{1}{2}\sqrt{1- \xi^2}\Bigr) \; \text{with} \;  \xi = \frac{1-\cos(\beta)}{3+\cos(\beta)} \;\;.
\end{equation}
Hence whenever $\beta=\pi\mod 2\pi$ the entangling operation is maximally entangling. Roughly, shorter entangling operations achieve more entanglement as they are less affected by dephasing. Hence we define the optimal velocity $v_{\text{opt}}$ for our scheme to be the fastest velocity for which the mean EF reaches a maximum i.e.\ $\beta(v_{\text{opt}})=\pi$. In Fig.~\ref{fig:flying-EF} we plot the EFs and probabilities around the optimal velocity $v_{\text{opt}}$. 

In summary, for this Letter we present an entanglement operation that is based on sensing small magnetic fields. Our calculations establish that this operation is fast enough to neglect dephasing for robust qubits. Moreover, imperfections in the scheme caused by vibrations of the NV centre are tiny for high frequencies and low amplitudes. Eventually remote qubits can be entangled by physically moving the NV centre mounted, for example, on an AFM-tip between the two qubits -- making the entangling operation of high value for a distributed quantum computer architecture.

\begin{acknowledgments}
\textit{Acknowledgements -} We thank Erik Gauger and Brendon Lovett for discussions. This work was supported by the DAAD, Linacre College, the National Research Foundation and Ministry of Education, Singapore, and the Japanese Ministry of Education, Culture, Sports, Science and Technology.
\end{acknowledgments}


\begin{thebibliography}{23}%
\makeatletter
\providecommand \@ifxundefined [1]{%
 \@ifx{#1\undefined}
}%
\providecommand \@ifnum [1]{%
 \ifnum #1\expandafter \@firstoftwo
 \else \expandafter \@secondoftwo
 \fi
}%
\providecommand \@ifx [1]{%
 \ifx #1\expandafter \@firstoftwo
 \else \expandafter \@secondoftwo
 \fi
}%
\providecommand \natexlab [1]{#1}%
\providecommand \enquote  [1]{``#1''}%
\providecommand \bibnamefont  [1]{#1}%
\providecommand \bibfnamefont [1]{#1}%
\providecommand \citenamefont [1]{#1}%
\providecommand \href@noop [0]{\@secondoftwo}%
\providecommand \href [0]{\begingroup \@sanitize@url \@href}%
\providecommand \@href[1]{\@@startlink{#1}\@@href}%
\providecommand \@@href[1]{\endgroup#1\@@endlink}%
\providecommand \@sanitize@url [0]{\catcode `\\12\catcode `\$12\catcode
  `\&12\catcode `\#12\catcode `\^12\catcode `\_12\catcode `\%12\relax}%
\providecommand \@@startlink[1]{}%
\providecommand \@@endlink[0]{}%
\providecommand \url  [0]{\begingroup\@sanitize@url \@url }%
\providecommand \@url [1]{\endgroup\@href {#1}{\urlprefix }}%
\providecommand \urlprefix  [0]{URL }%
\providecommand \Eprint [0]{\href }%
\providecommand \doibase [0]{http://dx.doi.org/}%
\providecommand \selectlanguage [0]{\@gobble}%
\providecommand \bibinfo  [0]{\@secondoftwo}%
\providecommand \bibfield  [0]{\@secondoftwo}%
\providecommand \translation [1]{[#1]}%
\providecommand \BibitemOpen [0]{}%
\providecommand \bibitemStop [0]{}%
\providecommand \bibitemNoStop [0]{.\EOS\space}%
\providecommand \EOS [0]{\spacefactor3000\relax}%
\providecommand \BibitemShut  [1]{\csname bibitem#1\endcsname}%
\let\auto@bib@innerbib\@empty
\bibitem [{\citenamefont {Cirac}\ \emph {et~al.}(1999)\citenamefont {Cirac}
  \emph {et~al.}}]{Cirac.Ekert.ea1999Distributedquantumcomputation}%
  \BibitemOpen
  \bibfield  {author} {\bibinfo {author} {\bibfnamefont {J.~I.}\ \bibnamefont
  {Cirac}} \emph {et~al.},\ }\href {\doibase 10.1103/PhysRevA.59.4249}
  {\bibfield  {journal} {\bibinfo  {journal} {Phys. Rev. A}\ }\textbf {\bibinfo
  {volume} {59}},\ \bibinfo {pages} {4249} (\bibinfo {year}
  {1999})}\BibitemShut {NoStop}%
\bibitem [{\citenamefont {Lim}\ \emph {et~al.}(2005)\citenamefont {Lim},
  \citenamefont {Beige},\ and\ \citenamefont
  {Kwek}}]{Lim.Beige.ea2005Repeat-Until-SuccessLinearOptics}%
  \BibitemOpen
  \bibfield  {author} {\bibinfo {author} {\bibfnamefont {Y.~L.}\ \bibnamefont
  {Lim}}, \bibinfo {author} {\bibfnamefont {A.}~\bibnamefont {Beige}}, \ and\
  \bibinfo {author} {\bibfnamefont {L.~C.}\ \bibnamefont {Kwek}},\ }\href
  {\doibase 10.1103/PhysRevLett.95.030505} {\bibfield  {journal} {\bibinfo
  {journal} {Phys. Rev. Lett.}\ }\textbf {\bibinfo {volume} {95}},\ \bibinfo
  {pages} {030505} (\bibinfo {year} {2005})}\BibitemShut {NoStop}%
\bibitem [{\citenamefont {Benjamin}\ \emph {et~al.}(2006)\citenamefont
  {Benjamin} \emph
  {et~al.}}]{Benjamin.Browne.ea2006Brokeredgraph-statequantum}%
  \BibitemOpen
  \bibfield  {author} {\bibinfo {author} {\bibfnamefont {S.~C.}\ \bibnamefont
  {Benjamin}} \emph {et~al.},\ }\href {\doibase 10.1088/1367-2630/8/8/141}
  {\bibfield  {journal} {\bibinfo  {journal} {New J. Phys.}\ }\textbf {\bibinfo
  {volume} {8}},\ \bibinfo {pages} {141} (\bibinfo {year} {2006})}\BibitemShut
  {NoStop}%
\bibitem [{\citenamefont {Barrett}\ and\ \citenamefont
  {Kok}(2005)}]{Barrett.Kok2005Efficienthigh-fidelityquantum}%
  \BibitemOpen
  \bibfield  {author} {\bibinfo {author} {\bibfnamefont {S.~D.}\ \bibnamefont
  {Barrett}}\ and\ \bibinfo {author} {\bibfnamefont {P.}~\bibnamefont {Kok}},\
  }\href {\doibase 10.1103/PhysRevA.71.060310} {\bibfield  {journal} {\bibinfo
  {journal} {Phys. Rev. A}\ }\textbf {\bibinfo {volume} {71}},\ \bibinfo
  {pages} {060310} (\bibinfo {year} {2005})}\BibitemShut {NoStop}%
\bibitem [{\citenamefont {Wrachtrup}\ \emph {et~al.}(2001)\citenamefont
  {Wrachtrup}, \citenamefont {Kilin},\ and\ \citenamefont
  {Nizovtsev}}]{Wrachtrup.Kilin.ea2001Quantumcomputationusing}%
  \BibitemOpen
  \bibfield  {author} {\bibinfo {author} {\bibfnamefont {J.}~\bibnamefont
  {Wrachtrup}}, \bibinfo {author} {\bibfnamefont {S.~Y.}\ \bibnamefont
  {Kilin}}, \ and\ \bibinfo {author} {\bibfnamefont {A.~P.}\ \bibnamefont
  {Nizovtsev}},\ }\href {\doibase 10.1134/1.1405224} {\bibfield  {journal}
  {\bibinfo  {journal} {Opt. Spectrosc.}\ }\textbf {\bibinfo {volume} {91}},\
  \bibinfo {pages} {429} (\bibinfo {year} {2001})}\BibitemShut {NoStop}%
\bibitem [{\citenamefont {Jelezko}\ \emph {et~al.}(2004)\citenamefont {Jelezko}
  \emph {et~al.}}]{Jelezko.Gaebel.ea2004ObservationofCoherent}%
  \BibitemOpen
  \bibfield  {author} {\bibinfo {author} {\bibfnamefont {F.}~\bibnamefont
  {Jelezko}} \emph {et~al.},\ }\href {\doibase 10.1103/PhysRevLett.92.076401}
  {\bibfield  {journal} {\bibinfo  {journal} {Phys. Rev. Lett.}\ }\textbf
  {\bibinfo {volume} {92}},\ \bibinfo {pages} {076401} (\bibinfo {year}
  {2004})}\BibitemShut {NoStop}%
\bibitem [{\citenamefont {Gaebel}\ \emph {et~al.}(2006)\citenamefont {Gaebel}
  \emph {et~al.}}]{Gaebel.others2006Room-temperaturecoherentcoupling}%
  \BibitemOpen
  \bibfield  {author} {\bibinfo {author} {\bibfnamefont {T.}~\bibnamefont
  {Gaebel}} \emph {et~al.},\ }\href {\doibase 10.1038/nphys318} {\bibfield
  {journal} {\bibinfo  {journal} {Nature Phys.}\ }\textbf {\bibinfo {volume}
  {2}},\ \bibinfo {pages} {408} (\bibinfo {year} {2006})}\BibitemShut {NoStop}%
\bibitem [{\citenamefont {Dutt}\ \emph {et~al.}(2007)\citenamefont {Dutt} \emph
  {et~al.}}]{Dutt.others2007QuantumRegisterBased}%
  \BibitemOpen
  \bibfield  {author} {\bibinfo {author} {\bibfnamefont {M.~V.~G.}\
  \bibnamefont {Dutt}} \emph {et~al.},\ }\href {\doibase
  10.1126/science.1139831} {\bibfield  {journal} {\bibinfo  {journal}
  {Science}\ }\textbf {\bibinfo {volume} {316}},\ \bibinfo {pages} {1312}
  (\bibinfo {year} {2007})}\BibitemShut {NoStop}%
\bibitem [{\citenamefont {Neumann}\ \emph {et~al.}(2010)\citenamefont {Neumann}
  \emph {et~al.}}]{Neumann.others2010Single-ShotReadoutof}%
  \BibitemOpen
  \bibfield  {author} {\bibinfo {author} {\bibfnamefont {P.}~\bibnamefont
  {Neumann}} \emph {et~al.},\ }\href {\doibase 10.1126/science.1189075}
  {\bibfield  {journal} {\bibinfo  {journal} {Science}\ }\textbf {\bibinfo
  {volume} {329}},\ \bibinfo {pages} {542} (\bibinfo {year}
  {2010})}\BibitemShut {NoStop}%
\bibitem [{\citenamefont {Taylor}\ \emph {et~al.}(2008)\citenamefont {Taylor}
  \emph
  {et~al.}}]{Taylor.Cappellaro.ea2008High-sensitivitydiamondmagnetometer}%
  \BibitemOpen
  \bibfield  {author} {\bibinfo {author} {\bibfnamefont {J.~M.}\ \bibnamefont
  {Taylor}} \emph {et~al.},\ }\href {\doibase 10.1038/nphys1075} {\bibfield
  {journal} {\bibinfo  {journal} {Nature Phys.}\ }\textbf {\bibinfo {volume}
  {4}},\ \bibinfo {pages} {810} (\bibinfo {year} {2008})}\BibitemShut {NoStop}%
\bibitem [{\citenamefont {Maze}\ \emph {et~al.}(2008)\citenamefont {Maze} \emph
  {et~al.}}]{Maze.Stanwix.ea2008Nanoscalemagneticsensing}%
  \BibitemOpen
  \bibfield  {author} {\bibinfo {author} {\bibfnamefont {J.~R.}\ \bibnamefont
  {Maze}} \emph {et~al.},\ }\href {\doibase 10.1038/nature07279} {\bibfield
  {journal} {\bibinfo  {journal} {Nature}\ }\textbf {\bibinfo {volume} {455}},\
  \bibinfo {pages} {644} (\bibinfo {year} {2008})}\BibitemShut {NoStop}%
\bibitem [{\citenamefont {Degen}(2008)}]{Degen2008Scanningmagneticfield}%
  \BibitemOpen
  \bibfield  {author} {\bibinfo {author} {\bibfnamefont {C.~L.}\ \bibnamefont
  {Degen}},\ }\href {\doibase 10.1063/1.2943282} {\bibfield  {journal}
  {\bibinfo  {journal} {Appl. Phys. Lett.}\ }\textbf {\bibinfo {volume} {92}},\
  \bibinfo {pages} {243111} (\bibinfo {year} {2008})}\BibitemShut {NoStop}%
\bibitem [{\citenamefont {Balasubramanian}\ \emph {et~al.}(2008)\citenamefont
  {Balasubramanian} \emph
  {et~al.}}]{Balasubramanian.Chan.ea2008Nanoscaleimagingmagnetometry}%
  \BibitemOpen
  \bibfield  {author} {\bibinfo {author} {\bibfnamefont {G.}~\bibnamefont
  {Balasubramanian}} \emph {et~al.},\ }\href {\doibase 10.1038/nature07278}
  {\bibfield  {journal} {\bibinfo  {journal} {Nature}\ }\textbf {\bibinfo
  {volume} {455}},\ \bibinfo {pages} {648} (\bibinfo {year}
  {2008})}\BibitemShut {NoStop}%
\bibitem [{\citenamefont {Schaffry}\ \emph {et~al.}(2011)\citenamefont
  {Schaffry} \emph {et~al.}}]{Schaffry.others2011SpinAmplificationMagnetic}%
  \BibitemOpen
  \bibfield  {author} {\bibinfo {author} {\bibfnamefont {M.}~\bibnamefont
  {Schaffry}} \emph {et~al.},\ }\href@noop {} {\enquote {\bibinfo {title} {Spin
  amplification for magnetic sensors employing crystal defects},}\ } (\bibinfo
  {year} {2011}),\ \Eprint {http://arxiv.org/abs/1104.0214} {arXiv:1104.0214
  [cond-mat.mtrl-sci]} \BibitemShut {NoStop}%
\bibitem [{\citenamefont {Levitt}(2008)}]{Levitt2008SpinDynamics}%
  \BibitemOpen
  \bibfield  {author} {\bibinfo {author} {\bibfnamefont {M.~H.}\ \bibnamefont
  {Levitt}},\ }\href@noop {} {\emph {\bibinfo {title} {Spin Dynamics - Basics
  of Nuclear Magnetic Resonance}}},\ \bibinfo {edition} {2nd}\ ed.\ (\bibinfo
  {publisher} {John Wiley \& Sons},\ \bibinfo {year} {2008})\BibitemShut
  {NoStop}%
\bibitem [{\citenamefont {Breuer}\ and\ \citenamefont
  {Petruccione}(2002)}]{Breuer.Petruccione2002TheoryofOpen}%
  \BibitemOpen
  \bibfield  {author} {\bibinfo {author} {\bibfnamefont {H.~P.}\ \bibnamefont
  {Breuer}}\ and\ \bibinfo {author} {\bibfnamefont {F.}~\bibnamefont
  {Petruccione}},\ }\href@noop {} {\emph {\bibinfo {title} {The Theory of Open
  Quantum Systems}}}\ (\bibinfo  {publisher} {Oxford University Press},\
  \bibinfo {year} {2002})\BibitemShut {NoStop}%
\bibitem [{\citenamefont
  {Wootters}(1998)}]{Wootters1998EntanglementofFormation}%
  \BibitemOpen
  \bibfield  {author} {\bibinfo {author} {\bibfnamefont {W.~K.}\ \bibnamefont
  {Wootters}},\ }\href {\doibase 10.1103/PhysRevLett.80.2245} {\bibfield
  {journal} {\bibinfo  {journal} {Phys. Rev. Lett.}\ }\textbf {\bibinfo
  {volume} {80}},\ \bibinfo {pages} {2245} (\bibinfo {year}
  {1998})}\BibitemShut {NoStop}%
\bibitem [{\citenamefont {Balasubramanian}\ \emph {et~al.}(2009)\citenamefont
  {Balasubramanian} \emph
  {et~al.}}]{Balasubramanian.others2009Ultralongspincoherence}%
  \BibitemOpen
  \bibfield  {author} {\bibinfo {author} {\bibfnamefont {G.}~\bibnamefont
  {Balasubramanian}} \emph {et~al.},\ }\href {\doibase 10.1038/nmat2420}
  {\bibfield  {journal} {\bibinfo  {journal} {Nature Materials}\ }\textbf
  {\bibinfo {volume} {8}},\ \bibinfo {pages} {383} (\bibinfo {year}
  {2009})}\BibitemShut {NoStop}%
\bibitem [{\citenamefont {Naydenov}\ \emph {et~al.}(2011)\citenamefont
  {Naydenov} \emph {et~al.}}]{Naydenov.others2011Dynamicaldecouplingof}%
  \BibitemOpen
  \bibfield  {author} {\bibinfo {author} {\bibfnamefont {B.}~\bibnamefont
  {Naydenov}} \emph {et~al.},\ }\href {\doibase 10.1103/PhysRevB.83.081201}
  {\bibfield  {journal} {\bibinfo  {journal} {Phys. Rev. B}\ }\textbf {\bibinfo
  {volume} {83}},\ \bibinfo {pages} {081201} (\bibinfo {year}
  {2011})}\BibitemShut {NoStop}%
\bibitem [{\citenamefont {Ge}\ \emph {et~al.}(2008)\citenamefont {Ge} \emph
  {et~al.}}]{Ge.others2008Modelingspininteractions}%
  \BibitemOpen
  \bibfield  {author} {\bibinfo {author} {\bibfnamefont {L.}~\bibnamefont {Ge}}
  \emph {et~al.},\ }\href {\doibase 10.1103/PhysRevB.77.235416} {\bibfield
  {journal} {\bibinfo  {journal} {Phys. Rev. B}\ }\textbf {\bibinfo {volume}
  {77}},\ \bibinfo {pages} {235416} (\bibinfo {year} {2008})}\BibitemShut
  {NoStop}%
\bibitem [{\citenamefont {Morley}\ \emph {et~al.}(2005)\citenamefont {Morley}
  \emph {et~al.}}]{Morley.others2005Hyperfinestructureof}%
  \BibitemOpen
  \bibfield  {author} {\bibinfo {author} {\bibfnamefont {G.~W.}\ \bibnamefont
  {Morley}} \emph {et~al.},\ }\href {\doibase 10.1088/0957-4484/16/11/001}
  {\bibfield  {journal} {\bibinfo  {journal} {Nanotechnology}\ }\textbf
  {\bibinfo {volume} {16}},\ \bibinfo {pages} {2469} (\bibinfo {year}
  {2005})}\BibitemShut {NoStop}%
\bibitem [{\citenamefont {Brown}\ \emph {et~al.}(2010)\citenamefont {Brown}
  \emph {et~al.}}]{Brown.others2010Electronspincoherence}%
  \BibitemOpen
  \bibfield  {author} {\bibinfo {author} {\bibfnamefont {R.~M.}\ \bibnamefont
  {Brown}} \emph {et~al.},\ }\href {\doibase 10.1103/PhysRevB.82.033410}
  {\bibfield  {journal} {\bibinfo  {journal} {Phys. Rev. B}\ }\textbf {\bibinfo
  {volume} {82}},\ \bibinfo {pages} {033410} (\bibinfo {year}
  {2010})}\BibitemShut {NoStop}%
\bibitem [{\citenamefont {Yun}\ \emph {et~al.}(1999)\citenamefont {Yun} \emph
  {et~al.}}]{Yun.Park.ea1999Regiospecificorientationof}%
  \BibitemOpen
  \bibfield  {author} {\bibinfo {author} {\bibfnamefont {K.}~\bibnamefont
  {Yun}} \emph {et~al.},\ }\href {\doibase 10.1007/BF02931919} {\bibfield
  {journal} {\bibinfo  {journal} {Biotechnol. Bioprocess Eng.}\ }\textbf
  {\bibinfo {volume} {4}},\ \bibinfo {pages} {72} (\bibinfo {year}
  {1999})}\BibitemShut {NoStop}%
\end{thebibliography}

%

\end{document}